\documentclass[useAMS,senatbib]{mn2e}
\usepackage{graphics}
\usepackage{graphicx}
\usepackage{psfig}
\def\ltapprox{\raise 2pt \hbox {$<$} \kern-1.1em \lower 5pt \hbox {$\approx$}}
\def\ltsim{\raise 2pt \hbox {$<$} \kern-0.8em \lower 4pt \hbox {$\sim$}}
\def\gtsim{\raise 2pt \hbox {$>$} \kern-0.8em \lower 4pt \hbox {$\sim$}}

\title[An elusive radio halo in the merging cluster Abell 781?]{An elusive radio halo in the merging cluster Abell 781?}

\author[T. Venturi et al.]{T. Venturi,$^{1}$\thanks{E-mail:
tventuri@ira.inaf.it} 
G. Giacintucci$^{2,1}$, D. Dallacasa$^{3,1}$, G. Brunetti$^{1}$, 
R. Cassano$^{3,1}$, G. Macario$^{1}$, \and R. Athreya$^{4}$\\
$^1$ INAF - Istituto di Radioastronomia, via P. Gobetti 101,I-40129 Bologna, Italy\\
$^2$ Department of Astronomy, University of Maryland, College Park, MD 20742--2421, USA\\
$^3$ Dipartimento di Astronomia,Universita' di Bologna, via Ranzani 1, I-40127 Bologna, Italy\\
$^4$ Indian Institute of Science Education and Research, Central Tower, Sai Trinity 
Building, Sutarwadi Road, Pashan, Pune 411021, India}

\begin{document}

\date{Accepted 2011 April 1. Received 2011 March 31; in original form 2011 February 24}

\pagerange{\pageref{firstpage}--\pageref{lastpage}} \pubyear{2011}

\maketitle

\label{firstpage}

\begin{abstract}

Deep radio observations of the galaxy cluster Abell 781 have been carried 
out using  the Giant Metrewave 
Radio Telescope at 325 MHz and have been compared to previous 610 MHz
observations and to archival VLA 1.4 GHz data.
The radio emission from the cluster is dominated by a diffuse source 
located at the outskirts of the X--ray emission, which we 
tentatively classify as a radio relic. We detected residual diffuse 
emission at the cluster centre at the level of S$_{\rm 325~MHz}\sim$15--20 mJy.
Our analysis disagrees with Govoni et al. (2011), and on the basis 
of simple spectral considerations we do not support their claim of a radio 
halo with flux density of 20--30 mJy at 1.4 GHz.
Abell 781, a massive and merging cluster, is an intriguing case. 
Assuming that the residual emission is  
indicative of the presence of a radio halo barely detectable at our
sensitivity level, it could be a very steep spectrum source.

\end{abstract}


\section{Introduction}

Cluster major mergers are among the most energetic phenomena in the Universe.
They release a total energy of the order of $10^{63}-10^{64}$ erg,
and it is nowadays accepted that they are the key ingredient to explain 
the origin and rarity of radio halos in galaxy clusters: shocks and turbulence 
are generated during such energetic events, and they deeply 
affect the thermal and non--thermal properties of the intracluster
medium (ICM).

Radio halos are the signposts of the non--thermal components in galaxy 
clusters. They are diffuse radio sources, whose size and
morphology are similar to those of the underlying hot ICM (e.g. Ferrari et 
al. 2008, Cassano 2009 and Venturi 2011 for recent reviews).
Their spectrum (defined as S$\propto\nu^{-\alpha}$) is steep, with typical 
values of the spectral index $\alpha$ in the range 1.2--1.4. However, 
recent high--sensitivity low frequency imaging led to the discovery of 
radio halos with much steeper spectra (Venturi 2011), with
spectral index $\alpha \sim 1.8-2$ 
(e.g. A\,521, Brunetti et al. 2008, Dallacasa et al. 2009; A\,697 
Macario et al. 2010).

Combined radio and X--ray studies provide strong support to the 
idea that radio halos are found only in unrelaxed clusters. 
Buote \cite{buote01} first showed a correlation between the 1.4 GHz 
radio power of halos, P$_{\rm 1.4~GHz}$, and the dipole power ratio
P$_{\rm 1}$/P$_{\rm 0}$ in the hosting cluster; based on {\it Chandra}
temperature maps, Govoni et al. \cite{govoni04}
found evidence for merging activity in clusters with radio halos.
Venturi et al. (2008, hereinafter V08) showed that all radio halos in 
the GMRT (Giant Metrewave Radio Telescope) radio halo survey are located 
in clusters with signs of dynamical disturbances. 
\\
More recently, Cassano et al. (2010, hereinafter C10) carried out 
a quantitative 
analysis of the radio halo--cluster merger scenario. They used all clusters 
in the GMRT radio halo cluster sample  (Venturi et al. 2007, 
hereinafter V07, and V08) with available high quality {\it Chandra} images 
(a total of 32 clusters) to characterize the presence of substructures by
three different methods. They showed that clusters with and without radio halos 
are well segregated according to all parameters indicating substructure: 
radio halos are associated with clusters currently undergoing a 
merger, while clusters without radio halo are usually more ``relaxed''.
Four clusters, however, are noticeable outliers in the correlations, being
disturbed systems with no detectable radio halo at the sensitivity limit
of the 610 MHz GMRT survey (V07 and V08).
One of the outliers, Abell 781 (hereinafter A\,781), has been 
observed by us with the 
GMRT at 325 MHz in a low frequency follow--up project of the GMRT radio halo 
survey (Venturi et al. to be submitted). In this letter we report
on our study on A\,781 and discuss the results in the light of 
the scenario of merger--induced formation of radio halos.

We adopt the $\Lambda$CDM cosmology with H$_0$=70 km s$^{-1}$ Mpc$^{-1}$, 
$\Omega_m=0.3$ and $\Omega_{\Lambda}=0.7$. At the redshift of A\,781 
(z=0.2984), 1$^{\prime \prime}=4.404$ kpc.

\section{A\,781 and its radio emission}

A\,781 (RA$_{\rm J2000}=09^h20^m23.2^s$, 
DEC$_{\rm J2000}=+30^{\circ}26^{\prime}15^{\prime\prime}$, z=0.2984) is
known to host a diffuse radio source South--East of its centre, 
tentatively classified as a relic on the basis of GMRT 610 MHz
observations (V08). 

The X--ray luminosity of the cluster, reported in the NORAS survey,
is $L_{\rm [0.1-2.4 keV]}$= 4.6$\times$10$^{44}$ erg s$^{-1}$ 
(Boh\"ringer et al. 2000). This value is 3 times lower than in the BCS
catalogue (Ebeling et al. 1998 and 2000) -- our reference in 
the selection of the GMRT radio halo 
sample -- and it is in line with measurements based on 
recent Chandra observations (Maughan et al. 2008, Wittman et al. 2006), 
and derived from a shallow ROSAT HRI exposure 
(S. Ettori, private communication).
The X--ray brightness distribution is very complex, with multiple peaks 
in the central region, 
and a secondary south--eastern condensation at $\sim 7^{\prime}$, 
associated with the galaxy cluster CXOU J\,092053+302800 located at z=0.291 
(Geller et al. 2005).  The candidate radio relic is located at 
the border of the X--ray emission from A\,781, in the
direction of CXOU J\,092053+302800 (see Fig. 5 in V08).

\subsection{The radio observations}

The cluster was observed with the GMRT at 325 MHz in January 2007, as
part of a project devoted to a low frequency follow--up study
of all radio halos and relics belonging to the GMRT radio halo survey
(Venturi et al. to be submitted).
The GMRT is excellent for imaging diffuse extended emission in 
crowded fields, 
allowing accurate subtraction of individual sources to properly image 
diffuse large scale emission. 

A\,781 was observed for a total of 8 hours, using the upper and lower 
side band (USB and LSB, respectively), left and right polarization, for a 
total observing bandwidth of 32 MHz. 
The data were collected in spectral--line mode with 128 channels/band, 
leading to a spectral resolution of 125 kHz/channel. 
The USB and LSB datasets were calibrated and reduced  individually  using 
the NRAO Astronomical Image Processing System (AIPS) package. 
Beyond the normal flagging of bad baselines, antennas and time ranges, a 
very accurate editing was carried out to identify and remove those data 
affected by radio frequency interference (RFI). The bandpass calibration 
was performed  using the  flux density calibrator (3C\,147). Due to 
strong RFI in the LSB dataset, only the upper portion of the band was 
used in the final imaging.
Self--calibration and imaging were carried out using
the same approach described in Macario et al. \cite{macario10} for A\,697.
We estimate that amplitude residual calibration errors are 
of the order of $\sim$8\%.
\\
We produced images in the resolution range from 
11.6$^{\prime\prime}$$\times$9.2$^{\prime\prime}$ to 
40$^{\prime\prime}$$\times$37.0$^{\prime\prime}$, with a 1$\sigma$ noise
$\sim$ 0.15--0.40 mJy beam$^{-1}$ going from high to low resolution. 
We finally imaged the cluster at the resolution of  
61.6$^{\prime\prime}$$\times$51.9$^{\prime\prime}$ reaching the noise of 
1$\sigma\sim 1$ mJy beam$^{-1}$.% (see Section 2.3).

%%%%%%%  fig. 1 - A781

\begin{figure*}%[htbp]
\centering
\includegraphics[angle=0,scale=1.36]{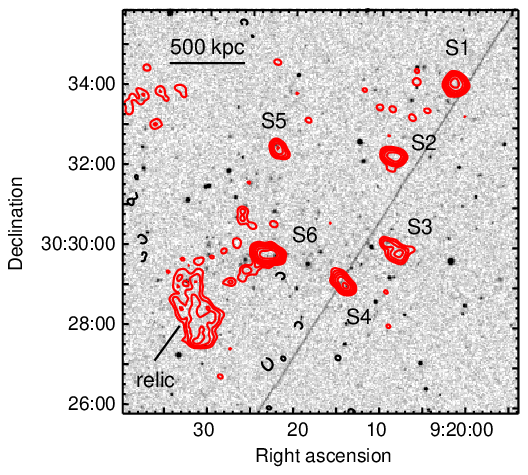}
\hspace{0.3cm}
\includegraphics[angle=0,scale=0.5]{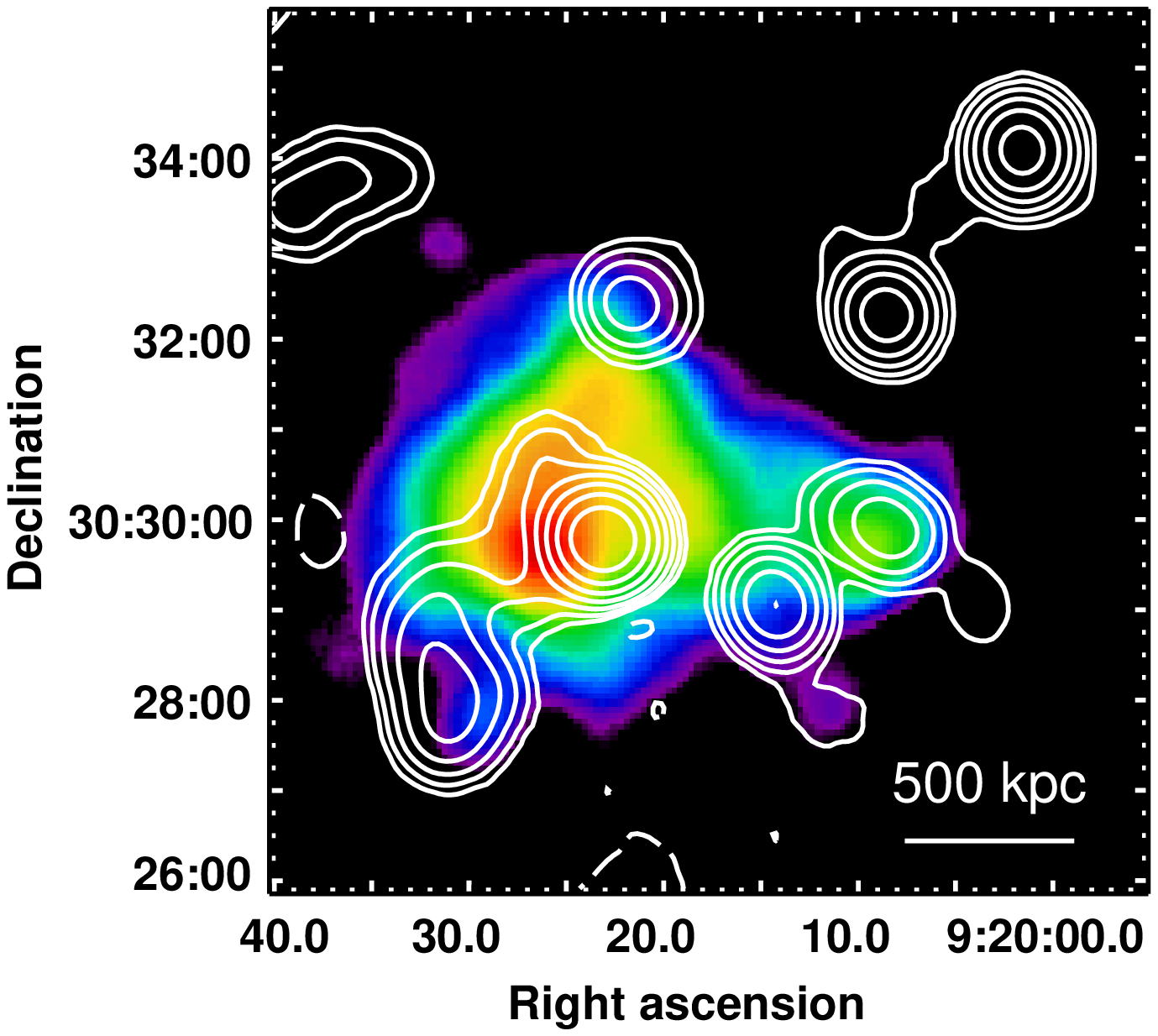}
\caption{Left -- GMRT 325 MHz radio contours of A\,781 at the resolution 
of 11.6$^{\prime\prime}\times9.2^{\prime\prime}$, position angle (p.a.) 
73$^{\circ}$, overlaid on the POSS--2 red plate. Contours (red positive,
black negative) start at $\pm$0.45 mJy beam$^{-1}$ (3$\sigma$) and scale 
by a factor of 2. Individual radio galaxies are
labelled from S1 to S6.
Right -- GMRT 325 MHz radio contours overlaid on the smoothed 
Chandra X--ray image. The resolution of the radio image is 
40$^{\prime\prime}$$\times$37$^{\prime\prime}$, p.a. $-81^{\circ}$.
Contours start at 1.2 mJy beam$^{-1}$ (3$\sigma$) and scale by a factor of 2.}
\label{fig:a781_tot}
\end{figure*}

%%%%%%%%% fig. 2 - A781 overlay 610-325 MHz
\begin{figure}%[htbp]
\centering
\includegraphics[angle=0,scale=1.40]{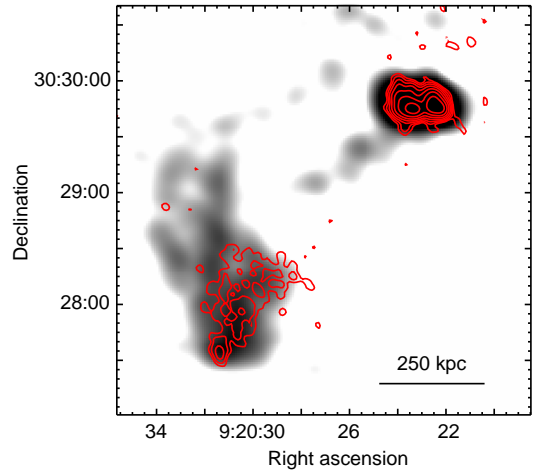}
\caption{GMRT 610 MHz contours in A\,781 overlaid on the GMRT 325 MHz
emission. The resolution of the 610 MHz image is 
6$^{\prime\prime}$$\times$5$^{\prime\prime}$, contours start at $\pm$0.25
mJy beam$^{-1}$
($\sim 3\sigma$) and are spaced by a factor of 2. The 325 MHz grey
scale image is the same as the left panel in Fig. \ref{fig:a781_tot}.}
\label{fig:a781_610_325}
\end{figure}
%%%%%%%%%%%%%%%%%%%%%%%%%%%%%%%%%%%%%%%%%%%%%%%%%%%%%%%%%

\subsection{The peripheral candidate relic}

Figure \ref{fig:a781_tot} shows the cluster radio emission at full and
low resolution, overlaid on the optical and X--ray emission respectively.
Beyond the individual sources, labelled S1 to S6
in the left panel (all resolved at this resolution), the candidate relic 
is the most striking feature in the A\,781 field.
As clear from Fig. \ref{fig:a781_610_325},
the source is much more extended at 325 MHz, compared to the 610 MHz image, 
with a largest angular size LAS $\sim 2^{\prime}$ (i.e. $\sim$ 790 kpc). 
Its radio brightness is peaked in the southernmost part of the source, 
which is also edge brightened, in agreement with the 610 MHz images. 
The flux density at 325 MHz is S$_{\rm 325~MHz}$=93.3$\pm$7.5 mJy, fairly 
consistent at all resolutions. The corresponding radio power is 
P$_{\rm 325~MHz}=$2.61$\times10^{25}$ W Hz$^{-1}$.
The flux density in the 610 MHz image does not increase even integrating
over the same extent imaged at 325 MHz. As mentioned in V08, the source is
well visible on the NVSS at 1.4 GHz. 

We looked into the NRAO VLA data archive in search for observations at higher 
frequencies. Two short observations exist, one at 1.477 GHz (C configuration, 
project AM469 observed on 15/03/1995), the second at 1.398 GHz 
(D configuration, project AO48 observed on 04/05/1984). We re--analyzed both
datasets, and 
the resulting L--band VLA images are shown in Fig. \ref{fig:a781_lp} overlaid 
on the 325 MHz cluster field. 
The flux density of the candidate relic is consistent 
with the measurement from the NVSS and amounts to 
S$_{\rm 1.4~GHz}$=15.3$\pm$0.5 mJy.
The spectral index in the range 325 MHz--1.4 GHz is $\alpha$=$1.25\pm0.06$.
The 610 MHz flux density (S$_{\rm 610~MHz}$=32$\pm$2 mJy, V08) is slightly
below the 325 MHz--1.4 GHz fit of the spectrum (by a factor of $\sim$20\%).
In V08 we discussed the missing flux for the diffuse radio 
emission of the 610 MHz GMRT Radio Halo Survey, and the  
missing flux in A\,781 is within the limits of our analysis.
We checked the flux density of 
the sources reported in Table 1 at 325 MHz, 610 MHz and 1.4 GHz (VLA--C array) 
using images of comparable resolution (of the order of 15$^{\prime\prime}$).
The flux density at 610 MHz is slightly understimated in all cases,
while the 325 MHz flux density 
values agree within 1$\sigma$ with those estimated from the WENSS 
(WEsterbork Northern Sky Survey, Rengelink et al. 1997) image.
The spectral index between 325 MHz and 1.4 GHz is 
consistent with nuclear emission from active galactic nuclei, 
i.e. $\alpha_{\rm 325 MHz}^{\rm 1.4 GHz}$ in the range $0.6\div0.8$ except
for source S5, which has $\alpha_{\rm 325 MHz}^{\rm 1.4 GHz} \sim 1.4$.
\\
An accurate analysis to thoroughly correct the 610 MHz flux density values is 
however beyond the scope of the present paper,
and for this source we will assume $\alpha_{\rm 325~MHz}^{\rm 1.4~GHz}$=1.25.
We point out that this value is the average over the whole emission at
325 MHz, but the size of the emission at 1.4 GHz and at 610 MHz is 
considerably smaller. The spectral index in the common portion of the 
emission is $\alpha_{\rm 325~MHz}^{\rm 1.4~GHz} \sim1$, allowing for the 
difficulties in ``isolating'' the 325 MHz flux density, while for the
remaining part we estimate a lower limit 
$\alpha_{\rm 325~MHz}^{\rm 1.4~GHz} < 2$. Such sharp jump in the spectral
index is intriguing, and worth it of further investigation.
\\
The nature of this source remains uncertain. Its overall properties --  
morphology and linear size, location at the border of the X--ray emission 
in the direction of the secondary X--ray peak associated with  
CXOU\,J092053+302800, and the steep spectrum -- coupled with the lack
of an obvious optical counterpart, suggest that it might be 
a cluster radio relic (see also V08). 
However, the uneven distribution of the spectral index
and the southern edge brightening at all frequencies are fairly unusual for 
relics, and we cannot rule out other possibilities,
such as a tailed radio source.

%%%%%%%%%% fin qui

\subsection{Radio emission from the central cluster region}

We checked for a possible extension of the relic towards the centre of 
A\,781, and/or for a radio halo undetected at higher frequencies.
We carried out flux density measurements over the inner $\sim$ 1.5 Mpc
around the cluster centre on the low resolution images
(from $30^{\prime\prime}$$\times$30$^{\prime\prime}$ to
$62^{\prime\prime}$$\times$52$^{\prime\prime}$), and on a residual image 
obtained after subtraction of the clean components associated with the 
individual radio sources (left panel of Fig. 1 and Table 1). 
We imaged the ``subtracted'' u--v dataset using natural weighting to 
enhance any possible presence of diffuse emission. 
Our images are shown as colour scale and contours in Figure \ref{fig:a781_lr}.
The residual image (Fig. \ref{fig:a781_lr}, right panel) clearly shows 
that the relic extends towards the centre of A\,781, in the direction of 
the double source S6.

The flux density measured in a central region of $\sim$ 1.5 Mpc in 
diameter, estimated by subtracting the contribution of the individual
sources (S4, S6 and the relic)
does not appreciably change with the resolution, and is of $\sim$ 15--20 mJy. 
Considering that the total flux density of S4, S6 
and the relic is 379.4 mJy (see Table 1), the residual flux density 
is $\sim$5\% of this value.
A similar value is found by integrating the residual image over the same 
1.5 Mpc region.
We thus consider this value an upper limit for the flux density of possible 
diffuse emission at the centre of A\,781, on a linear scale of 
the order of 1.5 Mpc.

\begin{figure*}%[htbp]
\centering
\includegraphics[angle=0,scale=0.41]{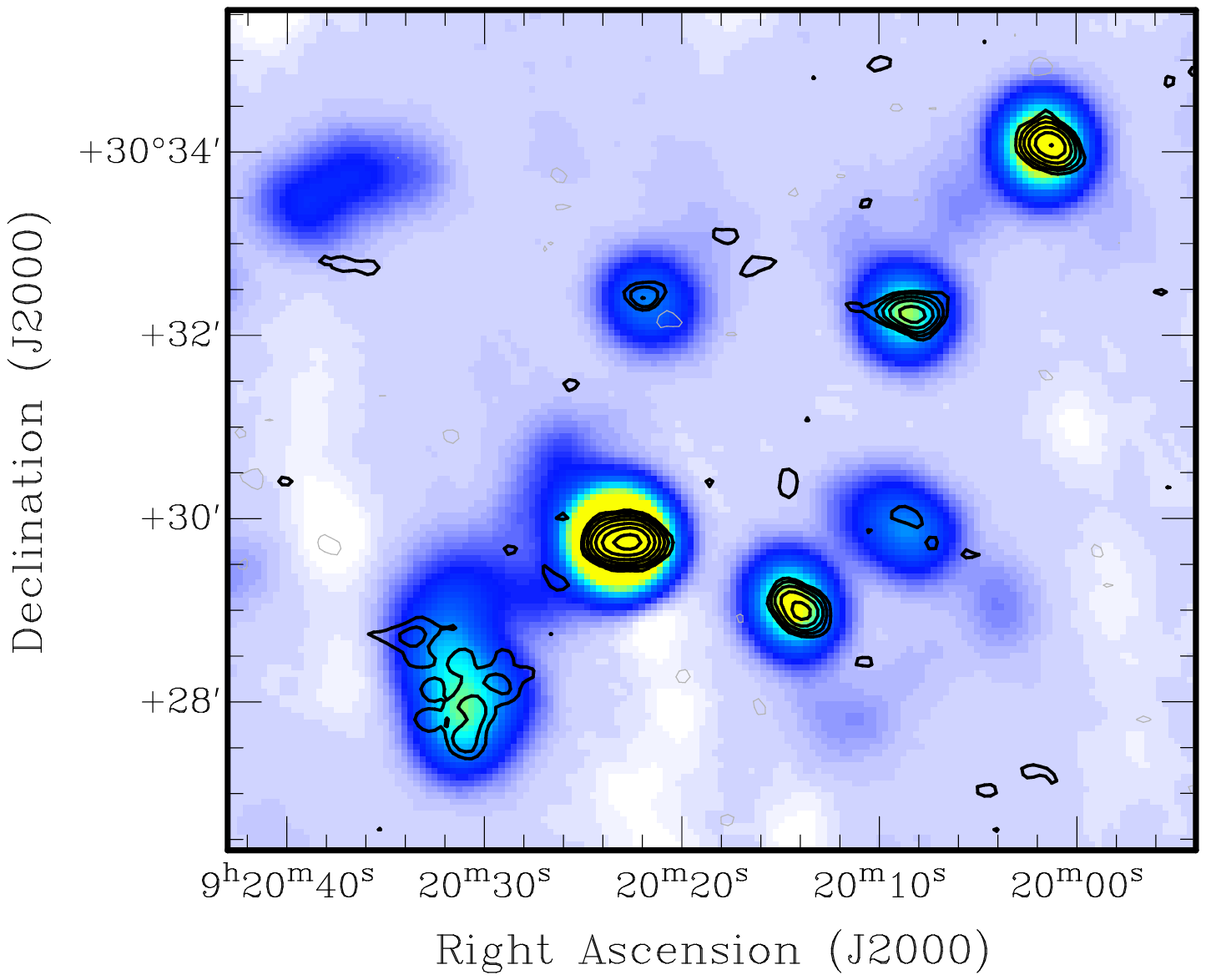}
\hspace{0.5cm}
\includegraphics[angle=0,scale=0.41]{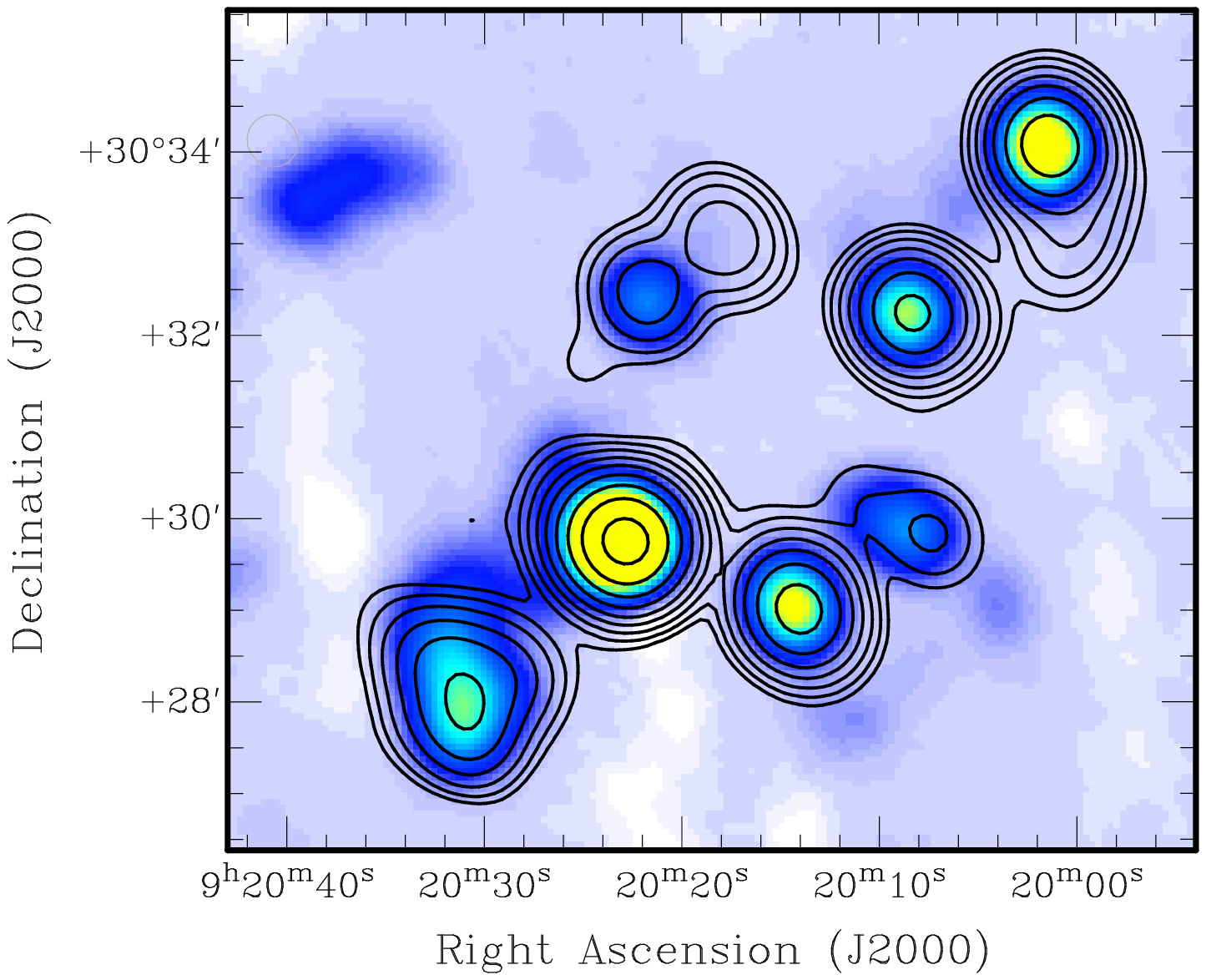}
\caption{Left -- VLA--C 1477 MHz contours at the resolution of
16.5$^{\prime\prime}$$\times$13.1$^{\prime\prime}$, in p.a. --86.4$^{\circ}$ 
overlaid on the 40$^{\prime\prime}$$\times$37$^{\prime\prime}$ 325 MHz GMRT image
(same as Fig. 1, right panel). Contours start at $\pm$0.45 mJy beam$^{-1}$
(3$\sigma$) and are spaced by a factor of 2.
Right -- VLA--D 1398 MHz contours of A\,781 at the resolution of
47.1$^{\prime\prime}$$\times$43.5$^{\prime\prime}$, in p.a. 18.1$^{\circ}$ 
overlaid on the same 325 MHz GMRT image. 
Contours start at $\pm$0.21 mJy beam$^{-1}$ (3$\sigma$, where $\sigma$ is
the confusion limit) and are spaced by a factor of 2.}
\label{fig:a781_lp}
\end{figure*}

\begin{figure*}%[htbp]
\centering
\includegraphics[angle=0,scale=0.41]{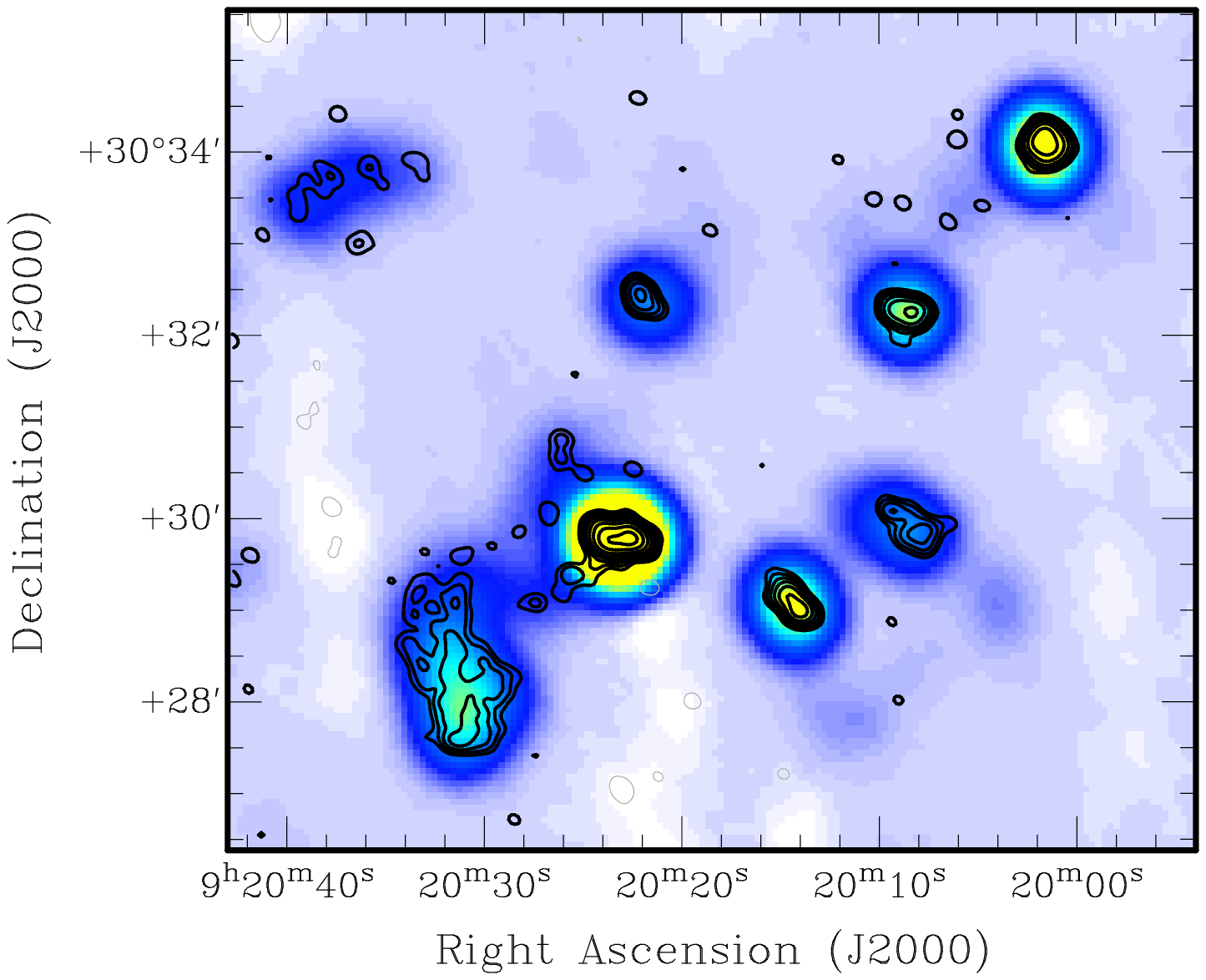}
\hspace{0.5cm}
\includegraphics[angle=0,scale=0.41]{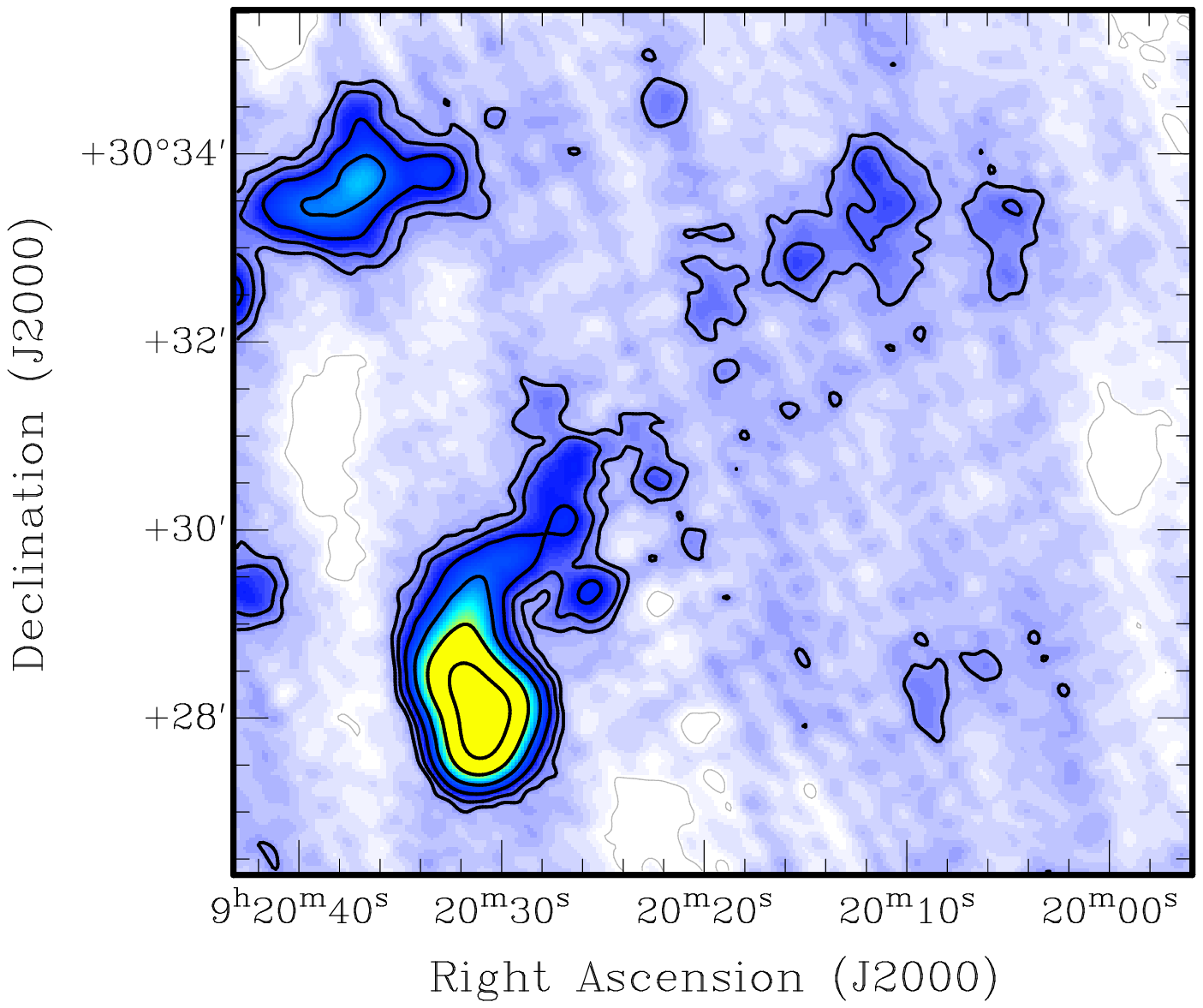}
\caption{Left -- GMRT 325 MHz image of A\,781 at the resolution of
40$^{\prime\prime}$$\times$37$^{\prime\prime}$, in p.a. 12.6$^{\circ}$ (blue
scale) with 11.6$^{\prime\prime}$$\times$9.2$^{\prime\prime}$ contours 
overplotted.
Contours are spaced by a factor of 2 starting from $\pm$0.45 mJy beam$^{-1}$.
Right -- GMRT 325 MHz residual image at the resolution of 
30$^{\prime\prime}$$\times$30$^{\prime\prime}$ with contours overlaid. 
Contours start from $\pm$0.5 mJy beam$^{-1}$ ($\sim 2\sigma$) and are 
spaced by a factor of 2.}
\label{fig:a781_lr}
\end{figure*}

The VLA L--band images we re--analyzed (Fig. \ref{fig:a781_lp}) 
are in agreement with our 325 MHz results. The different frequencies of the 
two observations do not allow an accurate subtraction of the individual 
sources from the VLA--C to the VLA--D dataset, since spectral effects would 
result in residual flux density  
from the individual sources subtracted in the 1.398 GHz VLA--D array image.
Moreover, the flatter spectrum of the individual sources compared to diffuse 
cluster sources (see Sect. 2.2 and Table 1) makes the source subtraction 
at 1.4 GHz more critical than at lower frequencies.
Hence we simply compared the sum of the flux density of the individual
sources S4, S6 and the relic to the total flux density
of the L--band images integrating over the whole area encompassing them.
No difference is detected in either images, and in both cases the values 
agree within $\ltsim$ 2\%.

The results of our analysis disagree 
with Govoni et al. \cite{govoni11}, who recently claimed the detection of a 
radio halo (with flux 20--30 mJy at 1.4 GHz) using the same VLA archival data.
They also derived an upper limit to the flux density  of the radio halo 
at 330 MHz ($\leq 135$ mJy) using an archive VLA observation, pointed 
1.5$^{\circ}$ away from the cluster centre, and give an upper limit of 
$\alpha_{\rm VLA~330~MHz}^{\rm VLA~1.4~GHz} \leq 1.3$ to its spectral slope. 
Our 325 MHz observations are about 5 times more sensitive than those
in Govoni et al., and rule out the presence of a halo in A\,781 
with flux density of 20--30 mJy at 1.4 GHz. 
The resulting spectral index, with our improved value of the 325 MHz residual
flux density, would be $\alpha_{\rm GMRT~325~MHz}^{\rm VLA~1.4~GHz} \leq 0.5$ 
(using a conservative limit, S$_{\rm 325 MHz}$$\leq$40 mJy) which is definitely 
unplausible for diffuse cluster sources.
We point out that one feature in the Govoni et al. residual image 
is the discrete source S3, which has an optical counterpart, as clear from 
the left panel of Fig. \ref{fig:a781_tot} and in V08; moreover, the
sources labelled C and D in their paper are clearly extended in the direction 
of the residual emission.

%%%%%%%%%% fig. 3 - A781 VLA over GMRT
%%%%%%%%%%%%%%%%%%%%%%%%%%%%%%%%%%%%%%%%%%%%%%%%%%%%%%%%%

%%%%%%%%%% fig. 4 - A781 low resolution and residuals

\section{Is there a radio halo in A\,781?}\label{sec:results}

A statistical connection between the dynamical state of massive clusters 
in the GMRT sample (V07 and V08) and the presence of radio halos has been 
derived by C10, 
confirming the picture where mergers switch--on radio halos in galaxy 
clusters.
A\, 781 is one of the few outliers in the Cassano et al. diagrams, lying in 
the region of dynamically disturbed clusters, but with no detected radio 
halo at 610 MHz (V08).

The observations at 325 MHz presented in this Letter do not allow a firm 
detection of a radio halo in the central region ($\sim$ 1.5 Mpc) of A\,781.
If we consider a spectral index $\alpha \sim 1.3$ between 325--1400 MHz,
the residual diffuse emission measured in our images, 
S$_{\rm 325~MHz}\sim$ 20 mJy, puts a conservative limit to the 1.4 GHz 
luminosity of a halo in A\,781 
P$_{\rm 1.4~GHz} \leq 6 \times (S_{0.3}/20)$$\times$10$^{23}$W~Hz$^{-1}$
(where $S_{0.3}$ is the flux at 325 MHz in mJy).

Our results do not challenge the cluster merger--radio halo
connection.
On the basis of the P$_{\rm 1.4~GHz}$--L$_{\rm X}$ correlation, the expected
1.4 GHz radio luminosity of the halo is still consistent with our upper
limit. As a matter of fact, the four known radio halos  with
radio power $\sim 10^{24}$ W~Hz$^{-1}$ hosted in clusters with X--ray 
luminosity 
L$_{\rm X} \leq 5$$\times$10$^{44}$erg~s$^{-1}$ (i.e. A\,2255, A\,2256, Coma and
A\,754) are only detected at low redshift, z$\leq$0.1.
On the other hand, if we assume the 1.4 GHz luminosity 
recently claimed by Govoni et al. \cite{govoni11}, the halo would lie about 
an order of magnitude above the radio/X--ray correlation. 
Incidentally, in their paper the halo is found consistent with that
correlation simply because the authors use
the old overestimated X--ray luminosity for A\,781, as given in the eBCS
catalogue (see Sect. 2).

Assuming that the residual emission at the centre of A\,781 does  
reveal the presence of an underlying very low brightness radio 
halo, the combination of 325 and 610 MHz provide further hints on its
spectral properties.
The 610 MHz upper limit and the residual emission at 325 MHz imply 
a spectral index of the emission steeper than $\alpha \geq$ 2.5.
Even accounting for the uncertainty in the 610 MHz upper limit,
which depends on the assumptions made on the unknown brightness 
distribution (see Brunetti et al. 2007 and V08), our experience shows 
that radio halos with 
S$_{\rm 610~MHz}$=10--20 mJy are well imaged in the GMRT Radio Halo
Survey (i.e. A\,697 at a similar redshift, V08). This suggests a
conservative upper limit of S$_{\rm 610~MHz}<10$ mJy, which would
still imply a steep spectrum halo, i.e. $\alpha$$>$1.5.
It is noteworthy that statistical expectations based on the 
turbulent re--acceleration model show that radio halos with steep
spectra should be quite common in merging clusters with masses
$\sim 10^{15}$ M$_{\odot}$,  similar to that of A\,781 (Cassano et al. 2006).

\section{Summary and Conclusions}\label{sec:summ}

We have presented deep GMRT 325 MHz observations of the unrelaxed and
luminous cluster A\,781, which is a noticeable outlier in the quantitative 
correlations connecting cluster mergers and the presence of a radio halo
(C10).
Our images show that the peripheral diffuse source is the dominant 
radio feature of the cluster, and only residual emission  at the level 
of S$_{\rm 325~MHz}$$\sim$ 15--20 mJy is found in a region of 
$\sim 1.5$ Mpc around the cluster centre (implying a conservative 
flux density limit of S$_{\rm 325~MHz}$$\sim$ 30--40 mJy).
This value improves the upper limit given in Govoni et al. \cite{govoni11}
by almost a factor of 5, and rules out the claim of a detection of
a radio halo at 1.4 GHz (Govoni et al. 2011) on the basis of 
simple spectral considerations.

With our data we cannot confirm the presence of a radio halo at the
centre of A\,781. 
If the 325 MHz residual emission is real, then it might trace 
an underlying halo with steep spectrum.
Future high sensitivity observations at lower frequencies combined with 
deeper observations at 610 MHz will allow us to clarify the nature
of the residual emission.

%%%%%%%%%%%%%%%%%%%%%%%%%%% Begin Table 1 %%%%%%%%%%%%%%%%%%%%%%%%%

\begin{table}%[t]
\caption[]{GMRT 325 MHz flux density of the discrete radio sources and 
spectral properties.}
\begin{center}
\begin{tabular}{lccrrc}
\hline\noalign{\smallskip}
Source  & RA $^a$& DEC $^a$& S$_{\rm peak}$ &  S$_{\rm tot}^b$ 
& $^{\rm c}\alpha_{\rm 325}^{\rm 1477}$ \\
name    & (h,m,s) & ($^{\deg}, ^{\prime}, ^{\prime \prime}$) &
(mJy/b) & (mJy) & \\
\noalign{\smallskip}
\hline\noalign{\smallskip}
S1    & 09 20 01.6& +30 34 06&  27.3$\pm$2.2 &  73.7$\pm$5.9 & 0.75 \\
S2    & 09 20 08.3& +30 21 15&  17.2$\pm$1.4 &  41.6$\pm$3.3 & 0.58 \\
S3    & 09 20 07.9& +30 29 51&   5.3$\pm$0.4 &  23.2$\pm$1.9 &  -   \\ 
S4    & 09 20 14.1& +30 29 01&  23.8$\pm$1.9 &  50.8$\pm$4.1 & 0.77 \\ 
S5    & 09 20 22.1& +30 32 25&  8.9$\pm$0.7  &  16.6$\pm$1.3 & 1.41 \\
S6    & 09 20 22.7& +30 29 47& 77.8$\pm$6.2  & 225.3$\pm$18.0 & 0.77 \\
Relic & 09 20 31.2& +30 27 35&  5.7$\pm$0.4  &  93.3$\pm$7.5 & see text\\
\hline{\smallskip}
\end{tabular}
\end{center}
Flux density values taken from the image shown in the left
panel of Fig. 1 and 4 (11.6$^{\prime\prime}$$\times$9.2$^{\prime\prime}$) 
after primary beam correction.\\
$^a$: coordinates of the radio peak; 
$^b$: total flux density measured using the AIPS task {\tt TVSTAT};
$^{\rm c}$ Spectral index derived from flux density measurements 
at the resolution of $17^{\prime\prime}$$\times$11$^{\prime\prime}$ at 325 MHz
and of $16^{\prime\prime}$$\times$13$^{\prime\prime}$ at 1477.40 MHz.
\label{tab:sources}
\end{table}

\section*{Acknowledgments}
We thank S. Ettori for his help in the X--ray checks. 
GMRT is run by the National Centre for Radio Astrophysics of the Tata 
Institute of Fundamental Research. Partial support was provided
by the Chandra grant AR0-11017X, NASA contract
NAS8-39073 and the Smithsonian Institution. 
S.G. acknowledges support by NASA through Einstein Postdoctoral Fellowship
PF0--110071 awarded by the Chandra X--ray Center which is operated by the 
Smithsonian Astrophysical Observatory under contract NAS8--03060.
This work is partially supported by INAF and ASI--INAF under grants 
PRIN--INAF2007, PRIN--INAF2008 and I/088/06/0.

\label{lastpage}

\end{document}